\documentstyle[12pt]{article}
\begin{document}
\begin{center}
ADAPTIVE TIME-DELAY HYPERCHAOS SYNCHRONIZATION IN LASER 
DIODES SUBJECT TO OPTICAL FEEDBACK(S)\\
\end{center}
\begin{center}
E. M. Shahverdiev \footnote{Regular Associate with the Abdus Salam ICTP;
 corresponding author;e-mail: shahverdiev@physics.ab.az}, R.A.Nuriev \\
Institute of Physics, 370143 Baku,Azerbaijan\\
R.H.Hashimov\\
Azerbaijan Technical University, 370073 Baku,Azerbaijan\\
K. A. Shore\\
School of Informatics, University of Wales, Bangor, Dean Street, Bangor, LL57 1UT, Wales, UK\\
~\\
ABSTRACT\\
\end{center}
In this paper a proposal is made of an adaptive coupling function for achieving 
synchronization between two lasers subject to optical feedback.Such a control scheme 
requires knowledge of the systems' parameters. For the first time we demonstate that 
when these parameters are not available on-line parameter estimation can 
be applied. Generalization of the approach to the multi-feedback systems 
is also presented.\\
PACS number(s):05.45.Xt, 05.45.Vx, 42.55.Px, 42.65.Sf
~\\
1.INTRODUCTION\\
~\\
The seminal papers  by  Pecora and  Carroll [1 ] and Ott, C.Grebogi and J.A.Yorke [2] on 
chaos synchronization have stimulated a wide range of research activity : a recent 
comprehensive review of such work is found in the focussed issues on chaos control [3] 
and references therein. Application of chaos control theory can be found in secure 
communications,optimization of nonlinear system performance and modeling brain activity 
and  pattern recognition phenomena  [3]. A particular focus of the work being the 
development of secure optical communications systems based on control and synchronization 
of laser chaos  [4-7]. It has been shown [8]  that 
 security cannot be guaranteed  in a communications format using simple chaotic 
systems -  ie those with a  single positive  Lyapunov exponent. It is thus appreciated 
that to obtain reliable  communications systems attention should be directed at  
hyperchaotic systems -  ie those 
with two or  more positive Lyapunov exponents. It has been claimed previously that the 
number of driving variables needed for synchronization in case of hyperchaotic systems 
should be equal to the number of positive Lyapunov exponents [3]. However, such a 
requirement is  highly undesirable in communication applications, as most communication 
schemes use just one signal for transmission [3]. More recently it was argued  [9-10 ],  
that hyperchaos control is possible 
using fewer  driving variables  than the number of positive Lyapunov exponents [9], and 
indeed even with zero- driving variables  using the method of parameter change advocated 
in [10]. Moreover , it has been shown recently that hyperchaos control is possible with a 
single variable even in the case of time delay systems, when the number of positive 
Lyapunov exponents ,in principle, can be infinite [11]. This result is of  particular 
importance for  the use of external cavity laser diodes for chaotic optical communications 
[ 7]. In addition to applications in communications the  implications of the study of 
synchronization phenomenon in time-delayed systems can be considered as a special case of 
spatio-temporal chaos control. Time-delay systems are infinite-dimensional and more 
interestingly by changing the time-delay one can obtain different numbers of the positive 
Lyapunov exponents [11 ].\\
In [11] and [12] use is made of both uni-directional and bi-directional couplings between 
the master and slave time-delay systems. Then an estimate is made of the coupling strength, for the given coupling function, needed for the synchronization between the drive and 
response time-delay systems. Usually two dynamical systems are termed synchronized if the 
differerence  between their states converges to zero for $t \rightarrow \infty $ [1-2]. 
Recently [13-14], a generalization of this  concept was proposed, where two systems are 
termed as being synchronized if a functional relation exists between the states of both 
systems.\\
In this paper we propose a general adaptive coupling (linking) function needed for 
synchronization between two time-delay systems. The approach does not require the 
imposition of threshold restrictions on the coupling strength. \\
Laser systems with optical feedback are prominent representatives of time-delay systems 
[6-7, 15-16]. We thus also apply the proposed approach to the case of synchronization 
between two lasers subject to optical feedback(s). Our results show that 
one can use a time 
delay coupling function to accomplish synchronization between the laser systems. This 
synchronization method is different from that of [15, 17]. We argue that such a diversity 
allows for more flexibility in practical control problems.\\
Usually, in the context of nonlinear dynamical systems, the method of adaptive control 
applies a feedback loop in order to drive the system parameter (or parameters) to the 
values required so as to achieve a target state. This is achieved by adding  the evolution 
of the parameter(s) to the evolution dynamics of the dynamical systems [18-20]. Such a 
scheme is adaptive, because the parameters which determine the nature of the dynamics 
self-adjust or adapt themselves to yield the desired dynamics.\\
In this paper we use the term 'adaptive' in a slightly different sense.
The proposed method of chaos synchronization between two chaotic systems can also be 
interpreted as follows: we apply a control law to the process model to reach the 
reference model (desired or target state), which in principle can be entirely different 
from the process model not only due to parameter(s) mismatches, but also by structure 
and/or dynamics; in other words, the task is to design a control force and apply it to 
the process model to reach an entirely different targeted state. Such a scheme is also 
adaptive, as in the above procedure the linkage function depending on the nature of the  
systems' dynamics, and structure  adapts  itself to yield the desired dynamics [21]. By 
definition, the adaptive principle is remarkably robust and  efficient in generic  
nonlinear systems and may therefore be of immense utility in a large variety of practical 
control problems.\\
~\\
2. ADAPTIVE COUPLING FOR CHAOS SYNCHRONIZATION \\
~\\
Following [11-12,21] we write the time-delay system under consideration in the form:
\begin{center}
\hspace*{5cm}$$\hspace*{5.2cm}\frac{dx}{dt}=f(x, x_{\tau}),\hspace*{9cm}$$
\hspace*{5cm}$$\hspace*{5.2cm}\frac{dy}{dt}=g(y, y_{\tau}) + W (x,y),\hspace*{6.4cm}(1)$$
\end{center}
\noindent where $f$ and $g$ are arbitrary time delay functions such that the 
corresponding dynamics exhibit chaotic behavior;$x_{\tau}:=x(t-\tau)$, 
$y_{\tau}:=y(t-\tau)$; where $\tau$ is the time-delay; the term $W (x,y)$ in equation (1) 
is responsible for coupling
(linkage)between the master (driving) (first equation in (1))and slave (response) (second 
equation 
in (1)) systems. The relaxation terms proportional to $x$ and $y$, usually written 
separately 
in the right hand sides of the system (1), are incorporated into the f and g. In addition, 
one must  keep in mind that in general the dynamical variables and $f, g$ and $W$ can be 
high dimensional vectors.\\
Let us choose the adaptive coupling function (or control input) $W(x,y)$ in the system (1) 
as:
\hspace*{4cm}$$\hspace*{4.5cm}W(x,y)=f(x, x_{\tau})-g(y, y_{\tau})-Q(x,y),\hspace*{4.5cm}(2)$$
with $Q(x,y)$ such that error dynamics of $x-y=e$ will be stable. For $Q(x,y)=B(x-y)=Be$ 
with negative $B$ we obtain:
\begin{center}
\hspace*{7cm}$$\hspace*{4.6cm}\frac{de}{dt}=Be, \hspace*{9.7cm}(3)$$
\end{center}
whose solution decays exponentially. Generalization to  the case of different delay 
functions 
with different time delays $\tau_{1}$ and $\tau_{2}$ is straightforward. In this case  
one can choose the coupling function
\hspace*{3cm}$$\hspace*{4.5cm}W(x, y)=f(x, x_{\tau_{1}})-g(y, y_{\tau_{2}})- Q(x,y).\hspace*{4.5cm}(4)$$
The idea behind the way of choosing the adaptive coupling (or control input) to achieve 
synchronization is to cancel the nonlinear terms of the system. We want to linearize the 
system to make it more tractable and to use linear control theory. Sometimes this approach 
is 
called  feedback linearization, for more details and pitfalls, see e.g. [22] and 
references therein.\\
As indicated above many laser communication systems are prominent representatives 
of  time-delay systems. In this work we apply the proposed approach to the case of 
synchronization between two semiconductor lasers subject to optical feedback. There can 
be different types of couplings between the slave and master systems.For example in [15] 
the light that is injected into the slave system is included in the equations in a way 
similar to the light coming from the external resonator.This approach is widely used to 
describe the effects of coherent light injection into semiconductor lasers.In this paper 
we propose a new type of coupling between master and slave systems to achieve 
synchronization between these systems. A general form for synchronization condition is 
obtained from a consideration of the following systems of the Lang-Kobayashi  equations 
[15,17] for the real electric field amplitude E(t), slowly 
varying phase $\Phi (t)$ and the  carrier number n(t) for the:master (with subscript M),
\hspace*{3cm}$$\hspace*{2cm}\frac{dE_{M}}{dt}=\frac{1}{2}G n_{M}E_{M}
 + k_{M}E_{M}(t-\tau )\cos (\omega_{0}\tau +\Phi_{M}(t)-\Phi_{M}(t-\tau)),\hspace*{2.5cm}$$
\hspace{2.5cm}$$\hspace*{2cm}\frac{d\Phi_{M}}{dt}=\frac{1}{2}\alpha Gn_{M}
-k_{M}\frac{E_{M}(t-\tau )}{E_{M}(t)}\sin (\omega_{0}\tau +
\Phi_{M}(t)-\Phi_{M}(t-\tau)),\hspace*{2.7cm}$$
\hspace*{4cm}$$\hspace*{2cm}\frac{dn_{M}}{dt}=(p-1)J_{th}-\gamma n_{M}(t)-(\Gamma +
Gn_{M})E_{M}^{2},\hspace*{5.7cm}(5)$$
and slave lasers (with subscript S),
\hspace*{2.5cm}$$\hspace*{2cm}\frac{dE_{S}}{dt}=\frac{1}{2}G n_{S}E_{S}
 + k_{S}E_{S}(t-\tau )\cos (\omega_{0}\tau +\Phi_{S}(t)-\Phi_{S}(t-\tau)) + W,\hspace{2.5cm} $$
$$\hspace*{2cm}\frac{d\Phi_{S}}{dt}=\frac{1}{2}\alpha Gn_{S} -k_{S}\frac{E_{S}(t-\tau )}{E_{S}(t)}\sin (\omega_{0}\tau +\Phi_{S}(t)-\Phi_{S}(t-\tau)),\hspace{3.5cm}$$
\hspace*{2.5cm}$$\hspace*{2cm}\frac{dn_{S}}{dt}=(p-1)J_{th}-\gamma 
n_{S}(t)-(\Gamma +Gn_{S})E_{S}^{2},\hspace*{6cm}(6)$$
coupled by the linkage function
$$W=K_{W}(E_{M}-E_{S})+\frac{1}{2}G (n_{M}E_{M}-n_{S}E_{S})
+k_{M}E_{M}(t-\tau )\cos (\omega_{0}\tau + \Phi_{M}(t)-\Phi_{M}(t-\tau))$$
\hspace*{3cm}$$\hspace*{2cm}-k_{S}E_{S}(t-\tau )\cos (\omega_{0}\tau +
\Phi_{S}(t)-\Phi_{S}(t-\tau)),\hspace*{6cm}(7)$$
where
$G$ is the differential optical gain;$\tau$ is the master laser's external cavity 
round-trip time;$\alpha$-the linewidth enhancement factor;$\gamma$- the carrier density 
rate;$\Gamma$-the cavity decay rate;$p$-the pump current relative to the threshold value 
$J_{th}$ of the solitary laser;$\omega_{0}$ is the angular frequency of the solitary laser;
$k$ is the feedback rate;$K_{W}$ is the coefficent determining the speed of achieving 
synchronization between the master and slave lasers.\\
One can see easily that  for the type of coupling with positive $K_{W}$ that the 
difference signal $e_{E}=E_{M}-E_{S}$ approaches zero, as the error dynamics in this case 
obey the following equation:
$$\hspace*{8.1cm}\frac{de_{E}}{dt}=-K_{W}e_{E}.\hspace*{5.5cm}(8)$$
(Throughout this paper we introduce the relaxation or damping term to overcome the 
necessity for  identical initial conditions in the coupled master and slave laser systems.)\\
In the above scheme of synchronization the master and slave systems' parameter,
namely the gain was the same for both systems. Generalization of the coupling function 
to the case of laser systems with different parameters is straightforward; for example, 
with different gain parameters the coupling function is:
$$W=K_{W}(E_{M}-E_{S})-\frac{1}{2}(G_{M}n_{M}E_{M} - G_{S}n_{S}E_{S})
 -k_{M}E_{M}(t-\tau )\cos (\omega_{0}\tau + \Phi_{M}(t)-\Phi_{M}(t-\tau))$$
\hspace*{12cm}$$\hspace*{3.1cm}+k_{S}E_{S}(t-\tau )\cos (\omega_{0}\tau +
\Phi_{S}(t)-\Phi_{S}(t-\tau)).\hspace*{5.4cm}(9)$$
As was pointed out in [23], in many representative cases, chaos synchronization can be 
understood from the existence of a global Lyapunov function of the difference signals. 
In other words, the global asymptotic stability can be investigated by the Lyapunov 
function 
approach [22]. For error dynamics $e_{E}$ (8), one can use the Lyapunov function
\hspace*{8cm}$$\hspace*{3.1cm}L=e_{E}^{2}.\hspace*{11.9cm}(10)$$
As 
\hspace*{8cm}$$\hspace*{3cm}\frac{dL}{dt}=-K_{W}e_{E}^{2},\hspace*{10.8 cm}(11)$$ 
can be made strictly negative for positive $K_{W}$ (except for $e_{E}=0$) we conclude 
that the asymptotic stability is global.\\
Thus based on our recent results we have several possibilities for achieving 
synchronization between chaotic laser diodes:according to [17] if the coupling between 
master and slave systems is of the form
\hspace*{4cm}$$\hspace*{3cm}W=\sigma E_{M}(t-\tau_{c}) \cos (\omega_{0}\tau_{c} + 
\Phi_{S}(t)-\Phi_{M}(t-\tau_{c})), \hspace*{4.7cm}(12)$$
(where $\sigma$ is the coupling strength between the master and slave lasers;
$\tau_{c}$ is the light propagation time from the right facet of the master laser to the 
right facet of the slave laser) then the synchronization condition is 
\hspace*{6cm}$$\hspace*{3cm}k_{M} = k_{S} + \sigma .\hspace*{11.3cm} (13) $$
In this paper we have proposed another type of linkage function (7) for the 
synchronization purposes without strict condition on the systems'  parameters.\\
\indent Multi-feedback and multi-delay systems are ubiquitous in nature and technology. 
Prominent examples can be found in biological and biomedical systems,laser physics,
integrated communications [24]. In laser physics such a situation arises in lasers subject 
to two or more optical or elctro-optical feedback. Second optical feedback could be useful 
to stabilize laser intensity [25]. Chaotic behaviour of laser systems with two optical 
feedback mechanism is studied in recent works [26]. To the best of our knowledge chaos 
synchronization between the multi-feedback systems is to be investigated yet. Having in 
mind enormous application implications of chaos synchronization e.g. in secure 
communication, investigation of synchronization in multi-feedback systems is of immense  
importance. It is well known that laser arrays hold great promise for space communication 
applications, which require compact sources with high optical intensities. The most 
efficient result can be achieved when the array elements are synchronized [27].\\
In the paper we only briefly condiser the case of adaptive synchronization in semiconductor 
lasers with double feedback. ( More detailed results on synchrinization regimes in 
the chaotic nonlinear systems will be presented elsewhere.) In the case of double feedback 
in the semiconductor lasers to the right-hand sides of the first equations (5) and (6) one 
has to add the terms 
$k_{M1}E_{M}(t-\tau_{1})\cos (\omega_{0}\tau_{1} +\Phi_{M}(t)-\Phi_{M}(t-\tau_{1}))$ and 
$k_{S1}E_{S}(t-\tau_{1})\cos (\omega_{0}\tau_{1} +\Phi_{S}(t)-\Phi_{S}(t-\tau_{1}))$.
(Of course corresponding terms should be added to the phase equations in (5) and (6).) 
Here $k_{M1,S1}$ are the feedback rate from the second mirrors in the master and slave 
lasers, respectively;$\tau_{1}$ is round trip time in the lasers' second external cavity. 
With this the linkage function to achieve adaptive synchronization between 
systems (5) 
and (6) will be written as follows:
$\hspace*{3cm}W= K_{W}(E_{M}-E_{S})+\frac{1}{2}G(n_{M}E_{M}-n_{S}E_{S}) +
k_{M}E_{M}(t-\tau)\cos(\omega_{0}\tau+\Phi_{M}(t)-\Phi_{M}(t-\tau))- 
k_{S} E_{S}(t-\tau) \cos (\omega_{0}\tau + 
\Phi_{S}(t)-\Phi_{S}(t-\tau)) + k_{M1}E_{M}(t-\tau_{1})\cos 
(\omega_{0}\tau_{1} +
\Phi_{M}(t)-\Phi_{M}(t-\tau_{1}))-k_{S1}E_{S}(t-\tau_{1})\cos (\omega_{0}\tau_{1} +
\Phi_{S}(t)-\Phi_{S}(t-\tau_{1}))$.
\begin{center}
3. ADAPTIVE SYNCHRONIZATION WITH UNKNOWN PARAMETERS\\
\end{center}
The systems parameters from eqs.(5-6) are required for the adaptive synchronization 
coupling function. In the case that these parameters are not available, one can apply 
the on-line parameter estimation method. In principle the number of unavailable parameters 
can be equal to the total number of systems' parameters. First in this paper we demonstrate  
the case of single parameter estimation, namely to gain estimation. Next we apply the 
approach to the case of double parameters estimation.
So let us suppose that the gain's estimated value  $G_{1}$  is different from the gain 
value $G$ required for synchronization. With the estimated value of gain the adaptive 
coupling function would be of the form
$$W=K_{W}(E_{M}-E_{S})+\frac{1}{2}G_{1}(n_{M}E_{M}-n_{S}E_{S})
+k_{M}E_{M}(t-\tau )\cos (\omega_{0}\tau + \Phi_{M}(t)-\Phi_{M}(t-\tau))$$
\hspace*{5cm}$$\hspace*{4.3cm}-k_{S}E_{S}(t-\tau )\cos (\omega_{0}\tau +
\Phi_{S}(t)-\Phi_{S}(t-\tau)),\hspace*{4.2cm}(14)$$
Under these conditions it is easy to verify that error dynamics now will satisfy the 
following equation:
\hspace*{3cm}$$\hspace*{3.8cm}\frac{de}{dt}=-K_{W} e + \frac{1}{2}(n_{M}E_{M}-n_{S}E_{S}) (G-G_{1}),\hspace*{4.7cm}(15)$$
In other words, for $G\neq G_{1}$ the error $e$ will not approach zero, as required for 
synchronization purposes. The situation can be rectified, if we add the following equation 
for the parameter estimation error $e_{G}=G-G_{1}$ to the previous equation (15):
\hspace*{5cm}$$\hspace*{3.8cm}\frac{de_{G}}{dt}=-e\frac{1}{2}(n_{M}E_{M}-n_{S}E_{S}) \hspace*{7.5cm}(16)$$
Now we shall demonstrate the the origin of the systems (15)-(16) is asymptotically stable,
i.e. that the synchronized state is asymptotically stable.
Indeed, by choosing the following Lyapunov function:
\hspace*{5cm}$$\hspace*{3.8cm}L=\frac{1}{2}(e^{2}+e_{G}^{2}),\hspace*{9.9cm}(17)$$
it is trivial to check that 
$$\hspace*{1.7cm}\frac{dL}{dt}=-K_{W}e^{2}+\frac{1}{2}(n_{M}E_{M}-n_{S}E_{S})e_{G}e
- \frac{1}{2}(n_{M}E_{M}-n_{S}E_{S})e_{G}e =-K_{W}e^{2}<0. \hspace*{0.8cm} (18)$$
Thus we demonstrate that the stability of the adaptive control law with unknown parameters 
is asymptotic. Next we suppose that apart from the gain, the master laser's estimated 
feedback rate value $k_{Mn}$ is different from the value of $k_{M}$ required for 
synchronization. Then with the estimated values of gain and feedback rate the adaptive 
coupling  function would be:
$W=K_{W}(E_{M}-E_{S})+\frac{1}{2}G_{1}(n_{M}E_{M}-n_{S}E_{S})
+k_{Mn}E_{M}(t-\tau )\cos (\omega_{0}\tau + \Phi_{M}(t)-\Phi_{M}(t-\tau))$$
\hspace*{5cm}$$\hspace*{4.3cm}\\
-k_{S}E_{S}(t-\tau )\cos (\omega_{0}\tau +
\Phi_{S}(t)-\Phi_{S}(t-\tau))$. With this $W$ the error dynamics is of the 
form:
$\frac{de}{dt}=-K_{W} e + \frac{1}{2}(n_{M}E_{M}-n_{S}E_{S}) (G-G_{1}) +
(k_{M}-k_{Mn}) E_{M}(t-\tau )\cos (\omega_{0}\tau + \Phi_{M}(t)-\Phi_{M}(t-\tau))$.
So for $G\neq G_{1}$ and $k_{M}\neq k_{Mn}$ synchronization is not achieved as $e$ with 
time is not approaching zero. Synchronization will take place only if the parameters' 
estimation errors $e_{G}=G-G_{1}$ and $e_{k}=k_{M}-k_{Mn}$ obey the dynamics:
$\frac{de_{G}}{dt}=-e\frac{1}{2}(n_{M}E_{M}-n_{S}E_{S})$ and 
$\frac{de_{Mn}}{dt}=-eE_{M}(t-\tau )\cos (\omega_{0}\tau + \Phi_{M}(t)-\Phi_{M}(t-\tau))$.
This time it can be shown by using the Lyapunov 
function $L=\frac{1}{2}(e^{2}+e_{G}^{2}+e_{k}^{2})$. It is evident that approach also 
applicable in the case double (multi-feedback) systems.\\
\indent In the paper we only have presented the case of complete chaos synchronization. 
Expansion of the approach to the lag and anticipating synchronizations is straightforward.\\
The practical implementation of the proposed scheme can be based on the approaches 
developed in [28-30]. In these papers it has been shown that a scheme of  chaos control 
for external cavity laser diodes can be effected where a periodic state of  the dynamics 
is selected from the chaotic dynamics. The key to that approach is the utilisation of an 
error signal which defines the difference between the chaotic state and the targeted state. 
As the target state is 
approached the generated error signal reduces to zero. It was shown in those papers that 
optoelectronic feedback provides a straightforward means for generating the requiste error 
signal [24-26]. It is noted that the linkage function defined in the present work can be 
expressed as an error signal between the dynamics of the master and slave external cavity 
lasers. As the linkage function brings the dynamics of the two laser systems into 
synchronization the corresponding error signal will diminish to zero. One approach to 
the practical implementation of the synchronization of the present scheme would thus again 
be based on the use of the optoelectronic feedback.\\
In other words, for the practical realization of the synchronization scheme we essentially 
inject the amplified difference signal between the master and slave lasers' outputs to the 
slave laser.\\
Comparing our approach with other widely known methods we notice that in general the 
present synchronization procedure is different from that of [15,17] and we argue that it 
offers more flexibility in practical control problems.\\
~\\
4.CONCLUSION\\
~\\
In this paper we have shown how one can synchronize two chaotic time delay systems in the 
general case by choosing an appropriate delay adaptive coupling function. We apply the 
proposed approach to the case of synchronization between two semiconductor lasers subject 
to optical feedback. For the first time we also have demonstrated that when the  
parameters of the systems to be synchronized are not available, then on-line parameter 
estimation can be applied. Generalization of the approach to the case of 
multi-feedback systems is also presented.\\
~\\
5.ACKNOWLEDGEMENTS\\
~\\
Concluding parts of the work have been done at the Abdus Salam ICTP. E.M.Shahverdiev 
kindly acknowledges a very helpful discussions with Professor H.A.Cerdeira. 
E.M. Shahverdiev 
also acknowledges support from the UK Engineering and Physical Sciences 
Research Council grant GR/R22568/01 and the Abdus Salam ICTP Associate scheme.\\

\end{document}